\documentclass[sigconf]{acmart}
\usepackage{multirow}
\usepackage{endnotes}

\AtBeginDocument{%
  }

\setcopyright{none}
\renewcommand\footnotetextcopyrightpermission[1]{}
\begin{document}

\let\footnote=\endnote

\title{Augmenting Human-Centered Racial Covenant Detection and Georeferencing with Plug-and-Play NLP Pipelines}

\author{Jiyoon Pyo}
\authornote{Both authors contributed equally to this research.}
\email{pyo00005@umn.edu}
\orcid{0009-0000-8746-4411}
\affiliation{%
  \institution{University of Minnesota}
  \country{USA}
}

\author{Yuankun Jiao}
\authornotemark[1]
\email{jiao0052@umn.edu}
\orcid{0000-0002-4188-5066}
\affiliation{%
  \institution{University of Minnesota}
  \country{USA}
}

\author{Yao-Yi Chiang}
\orcid{0000-0002-8923-0130}
\email{yaoyi@umn.edu}
\affiliation{%
  \institution{University of Minnesota}
  \country{USA}
}

\author{Michael Corey}
\orcid{0009-0000-1455-9612}
\email{corey101@umn.edu}
\affiliation{%
  \institution{University of Minnesota}
  \country{USA}
}

\renewcommand{\shortauthors}{Pyo et al.}

\begin{abstract}

Though no longer legally enforceable, racial covenants in 20th-century property deeds continue to influence spatial and socioeconomic inequalities. Understanding this legacy requires identifying racially restrictive language and geolocating the associated properties. The Mapping Prejudice project addresses this by engaging volunteers using the Zooniverse crowdsourcing platform to transcribe racial covenants from scanned property deed images and using transcribed legal descriptions to link those covenants to modern parcel maps. Mapping Prejudice does not seek to fully automate racial covenants research, as it has found that crowdsourcing has powerful social benefits in addition to its technical superiority over fully automated methods. The project has historically relied on simple lexicon-based searching and, more recently, on fuzzy matching to flag suspected racial covenants for community mapmakers to review. However, the introduction of fuzzy matching has significantly increased the number of false positives that volunteers must review, negatively impacting their user experience and potentially presenting scalability challenges. Further, while many properties can be mapped automatically based on user transcription, others require time-consuming manual geolocation via human research.

To address these limitations, we present a human-centered computing strategy with two plug-and-play natural language processing (NLP) pipelines: (1) a context-aware text labeling model that flags racially restrictive language with high precision,\footnote{\url{https://github.com/knowledge-computing/MP-term-analyses}} and (2) a georeferencing module that extracts geographic descriptions from deed documents and resolves them to real-world geographic regions.\footnote{\url{https://github.com/knowledge-computing/MP-text-to-land}} When evaluated on historical deed documents from six counties in Minnesota and Wisconsin, our system reduces false positives in racial term detection by 25.96\% while maintaining a recall of 91.73\%, and achieves 85.58\% accuracy in georeferencing to correct location within 1×1 square-mile geocoordinate ranges. These tools can accelerate volunteer participation and manual cleanup, strengthening public engagement by improving document filtering and enriching spatial annotations.

\textit{Warning: This paper contains offensive racial terms from actual historical deed documents.}

\end{abstract}



\begin{teaserfigure}
  \includegraphics[width=\textwidth]{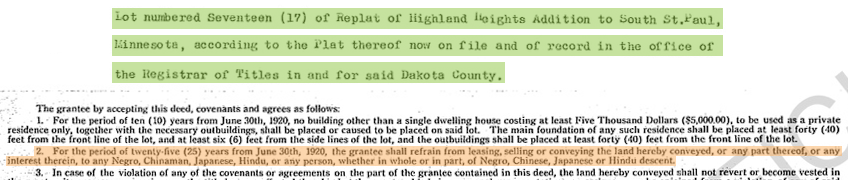}
  \caption{Deed from the year 1944, Dakota County, Minnesota. The section highlighted in orange contains discriminatory language; the section in green indicates the property's location.}
  \Description{Deed from year 1944, Dakota County, Minnesota. The section highlighted in orange contains discriminatory language; the section in green indicates the property's location. In the Mapping Prejudice project, volunteers transcribe the racially discriminatory language and summarize location information found in historical deed documents.}
  \label{fig:teaser}
\end{teaserfigure}

\received{23 July 2025}

\maketitle

\section{Introduction}

Racial covenants are clauses written into 20th-century property deeds that prohibited individuals who were deemed not to be White from buying, renting, or occupying properties in many neighborhoods~\cite{c:welsh2018}. Though the racial covenants were ruled to be unenforceable in \textit{Shelley v. Kraemer} (1948)~\cite{c:shelleyvkraemer} and outlawed by the Fair Housing Act of 1968, their historical legacies persist in contemporary patterns of segregation~\cite{c:sood2019}, wealth inequality~\cite{c:santucci2025}, and environmental injustice~\cite{c:walker2024}. For example, \citet{c:walker2024} shows that the distribution of racial covenants correlates with lower present-day temperatures and higher tree cover. Making this history visible is, therefore, essential to advancing public understanding of structural racism~\cite{c:mattke2022} and informing evidence-based urban policy~\cite{c:walker2023, c:goetz2024}.

The Mapping Prejudice\footnote{\url{https://mappingprejudice.umn.edu/}} project~\cite{c:mappingprejudice} is a community-engaged research project that addresses these needs by inviting ``community mapmaking'' volunteers to transcribe racially restrictive covenants and identify the location of the associated properties. More than 12,000 community mapmakers have participated in Mapping Prejudice transcription, identifying and mapping over 75,000 racially covenanted properties in several states~\cite{c:mappingprejudice}. Mapping Prejudice's public racial covenants data\footnote{\url{https://github.com/UMNLibraries/mp-us-racial-covenants}} has been used by educators, economists, geographers, and historians for secondary research and, most importantly, to drive real-world change~\cite{c:minneapolis2040, c:Worthington02012020, c:Mogush02012020,c:mccormick2020,c:kahlenberg2019,c:clemence2021}.

While other researchers have called on Mapping Prejudice to attempt to fully automate with artificial intelligence and dispense with its volunteer transcribers, the project has rejected this as antithetical to its theory of change, in addition to the profound technical limitations of a fully automated approach. The project team's experience has demonstrated that research seeking to power social change must actively involve community members outside the university. Personal interactions with these primary historical sources sit at the heart of Mapping Prejudice. These experiences provide people with a new frame for understanding their personal experiences and their communities, inspiring them to push for changes that dismantle structural racism~\cite{c:ehrman2022, c:mattke2022}. However, relying solely on manual transcription for all deed documents is inefficient, as many contain clauses unrelated to racial covenants. The massive volume of documents makes the process tedious for the volunteers. This primacy of the volunteer contribution also guides the direction of software development for Mapping Prejudice, where development decisions are calibrated to enhance the experience of the volunteers who transcribe these racist property records~\cite{c:mappingprejudice}.


To detect racially restrictive terms, the current approach of Mapping Prejudice utilizes a simple lexicon-based fuzzy-matching pipeline as an initial filter~\cite{c:mappingprejudice} to flag racist language across millions of pages of deeds and other property records. While this method reduces the volunteer workload, it still yields a large number of false positives and requires continual lexicon updates whenever new variants of racially restrictive terminology are identified. The high false positive rate negatively affects the volunteer experience during transcription, as it can lead to slower progress when repeatedly reviewing irrelevant or misleading results.

Identifying geographic regions likely to contain racially restrictive terms based on environmental features and spatial segregation patterns may also enhance the recall rate of racial covenant detection. Currently, the Mapping Prejudice project relies on volunteers to annotate geographic descriptions in scanned deed documents through the crowdsourcing platform. While many properties can automatically be linked to modern properties through this transcription, mapping properties with complicated legal descriptions is labor-intensive. Technologies that assist humans in recognizing and georeferencing the property legal descriptions within a confined area can streamline the linkage between deeds and actual locations.

To alleviate these bottlenecks, we present a human-centered strategy with two modular entity recognition pipelines that detect (1) racially restrictive clauses based on context evidence and (2) property geographic descriptions and georeference the deeds. Our preliminary experiment demonstrates a reduction in false positive rates by 25.96\% in racial terminology detection, while preserving high recall, allowing volunteers to focus on a subset of relevant historical deeds. For georeferencing, our pipeline achieves 86.39\% and 85.58\% accuracy at a 6×6 square-mile and 1×1 square-mile resolution, respectively. These pipelines aim to generate results that can work in conjunction with humans in crowd-sourcing settings, thereby reducing the potential burden of dealing with large numbers of documents and improving community engagement.

\section{Related Works}
We review two areas of related work most relevant to our study: research on racial bias in NLP (Section \ref{s:related-bias}) and approaches to georeferencing historical deed documents (Section \ref{s:related-georeference}).

\subsection{Racial Bias} \label{s:related-bias}
A large body of research focuses on identifying and mitigating racial and social biases in text corpora and language models. Prior studies investigate bias detection across various domains, using techniques such as counterfactual data augmentation~\cite{c:barriere2024}, adversarial training~\cite{c:zhang2018,c:sun2019}, and loss function modifications~\cite{c:qian2019,c:mamta2024} to reduce the impact of biased representations~\cite{c:bolukbasi2016}.

While our work also engages the concept of bias, specifically, allocational bias~\cite{c:blodgett2020}, it differs in both intent and application. Rather than mitigating bias for fairness in downstream tasks, we aim to leverage the presence of bias in historical deed documents to recover and expose the systemic patterns of racial exclusion encoded in the language of property records. Unlike prior work, our objective is to support public-facing historical accountability efforts by identifying and highlighting racially restrictive language.

\subsection{Georeferencing Historical Deed Documents} \label{s:related-georeference}
Georeferencing is the process of extracting location references from text or visual landmarks and translating them to geographic coordinates or boundaries~\cite{c:gritta2020}. With text-based georeferencing, this typically involves two steps: (1) recognizing toponyms (i.e., place names) in the text through methods such as named entity recognition (NER) and (2) linking the extracted toponyms to corresponding geolocations~\cite{c:aldana2020}. Georeferenced documents are used in various applications, including spatial humanities~\cite{c:hu2023}.

20th-century U.S. deed documents commonly reference two complementary land-survey systems: the Recorded Plat Survey System (RPSS) and the Public Land Survey System (PLSS)~\cite{c:reuterslot2025}. RPSS records properties using named subdivisions along with higher-level jurisdictions (i.e., State, County, and City). A subdivision is a legally platted division of land for sale or development, typically 300 to 7,000 square meters in area. Real estate entities and urban planning departments widely use RPSS in densely populated metropolitan and suburban areas~\cite{c:reuterslot2025}. PLSS overlays the United States with a rectilinear grid defined by Township, Range, and Section~\cite{c:MN_PLSS_2023, c:WI_PLSS_2019}. A Township represents either a 6-mile measurement north or south from a referenced baseline or a 6-mile-square parcel, while a Range represents a 6-mile measurement east or west from the principal meridian. In Minnesota and Wisconsin, the combination of Township and Range defines an approximately 6×6 square-mile grid block. Each grid block is further divided into 36 Section blocks, each measuring about 1×1 square-mile~\cite {c:plss2012, c:usgs2025plss}. 

Identifying references of both RPSS and PLSS in deed documents enables precise georeferencing, which is essential for studying patterns of racial segregation and environmental injustice. A system that supports georeferencing from deed text can enhance the manual transcription process by reducing repetitive work for volunteers, thereby optimizing the user experience and fostering stronger community engagement.

Off-the-shelf NLP libraries such as spaCy\footnote{https://spacy.io} can recognize broad geopolitical entities such as Countries, States, and Cities. However, they cannot resolve PLSS references or other property legal descriptions, which are essential for georeferencing historical deeds. To bridge this gap, we train a deed-specific named-entity recognizer that identifies the seven entities: State, County, City, Subdivision, Township, Range, and Section. These entities are subsequently fed into a resolver that leverages geographic boundaries to translate historical references into their corresponding coordinate ranges, thereby enabling granular analyses of historical deed documents.

\section{Method}
In the following sections, we present our methodological framework.

\subsection{Racial Entity Identification} \label{sec:racial-entity-identification}
To develop our racial entity identification pipeline, we use the data provided by Mapping Prejudice, comprising over 1.4 million historical property records pages from four counties: Anoka, Dakota, and Washington in Minnesota, and Milwaukee in Wisconsin. The fuzzy-matching system, which is currently integrated into the Mapping Prejudice team's pipeline, identifies 28,147 pages from these four counties as potential covenants. Through the crowdsourcing platform, volunteers annotate these flagged deeds at the document level, marking the presence of racially restrictive language.

To obtain token-level details from document-level annotations, we align the volunteer-flagged documents with optical character recognition (OCR) results and match each word against a hand-curated lexicon of 73 racially restrictive terms and phrases (e.g., ``Caucasian'', ``Hindu'', ``domestic servants''). This creates 147,555 labeled sentences from 7,244 documents, with 15\% of the tokens labeled as racially restrictive. To increase the diversity and generalizability of the training data, we adapt two stereotype corpora, StereoSet~\cite{c:nadeem2021} and CrowS-Pairs~\cite{c:nangia2020}. From these corpora, we extract sentences tagged with racial stereotypes and their associated target entities. This contributes an additional 3,362 labeled sentences with racially restrictive terms.

OCR noise is common in historical deeds, which can degrade the recall of fuzzy string matching. Moreover, relying on a predefined lexicon cannot capture paraphrases or novel phrasings that are missing from the predefined term list. To address both issues, we train a token-classification model so that the sentence-level context helps (1) recover distorted tokens introduced by OCR errors and (2) identify previously unseen but semantically equivalent restrictive terminologies. 

Therefore, we fine-tune ModernBERT~\cite{c:warner2024} for token classification. We choose ModernBERT over BERT~\cite{c:devlin2019} and its variants due to its ability to process longer context windows and its improved inference speed. The two characteristics are especially beneficial for parsing deed sentences since deed documents often exceed the context limit of standard BERT models. The fine-tuned model labels each word in the document as either part of a racially restrictive term or not.



\subsection{Georeferencing Racial Deeds} \label{s:georeferencing}
Our georeferencing pipeline consists of two steps: first, identifying geoentities in deed text, and then using Township, Range, and Section to resolve each deed to its modern location. We build our pipeline on Mapping Prejudice's publicly available racial covenants dataset, which includes 3,453 racially covenanted property deeds and 12,180 associated parcel polygons from six counties: Anoka, Dakota, Washington, Olmsted, and Sherburne in Minnesota, and Milwaukee in Wisconsin. For each deed, we utilize the OCR text, volunteer- or expert-labeled textual attributes (i.e., County, City, and Subdivision), and parcel geometries manually digitized by volunteers (i.e., polygons). We use the deduplicated deed documents from 70\%, 10\%, 20\% parcel geometries for training, validating, and testing, respectively. 

To create ground truth labels for named-entity recognition (NER), we assign geoentity-relevant words in the deed documents to one of the seven categories: State, County, City, Subdivision, Township, Range, and Section. We take State, County, City, and Subdivision (RPSS entities) directly from the volunteer- or expert-provided labels in the dataset. However, as volunteers and experts do not label Township, Range, and Section, we design regular expressions targeting their typical ``keyword + number'' format (e.g., ``Township numbered Eight (8) North'', ``Section 25''). To ensure the accuracy of the extracted components and to account for OCR errors and formatting variations, we manually review all detected instances of Township, Range, and Section to generate the final ground-truth labels. 

As there are no existing NER models tailored to property deeds, we train a spaCy v3 named-entity model from scratch to identify all seven target entities. Using the Township, Range, and Section entities, we normalize textual numerals to numeric values (e.g., ``Township One hundred and six North'' to ``106'') and create (Township, Range, Section) triples or (Township, Range) tuples that correspond to 1×1 and 6×6 square-mile resolutions. When identical identifiers occur across states, we disambiguate them with the Range orientation (i.e., ``West'' or ``East''). To map each deed's historical references to modern location, we resolve these triples and tuples using the PLSS geographic data for Minnesota~\cite{c:MN_PLSS_2023} and Wisconsin~\cite{c:WI_PLSS_2019}.

\section{Experimental Results and Analysis}
In the following sections, we present our experimental results and analysis for each pipeline.

\subsection{Racial Entity Identification} \label{s:racial-entity-result}


We report the performance on document-level flagging and token-level term tagging, to match Mapping Prejudice's needs to (1) find any covenant and (2) localize the terms. We compare the precision and recall of our approach with those of the fuzzy match baseline. As the volunteers review only flagged documents, we assume the recall score of the fuzzy match method as 100\%, though in practice, OCR errors do lead to an unknown number of false negatives.
We summarize the result in Table \ref{t:rei-results}.

\begin{table}[htbp]
    \centering \small
    \begin{tabular}{|c|c|c|c|c|}
        \hline
         & \multicolumn{2}{|c|}{\textbf{Per Token}  (\%)} & \multicolumn{2}{|c|}{\textbf{Per Document}  (\%)} \\
        \hline
        \textbf{Method} & \textbf{Recall} & \textbf{Precision} & \textbf{Recall} & \textbf{Precision} \\
        \hline \hline
        Fuzzy match & - & 67.73 & - & 59.03 \\
        \hline
        Proposed & 91.73 & \textbf{93.69} & 88.86 & \textbf{93.06} \\
        \hline
    \end{tabular}
    \caption{Token- and document-level precision and recall for fuzzy match baseline and our proposed pipeline.
        }
    \label{t:rei-results}
\end{table}

Our proposed model achieves a 25.96\% increase in precision at the token level and a 34.03\% increase at the document level, while maintaining strong recall (91.73\% and 88.86\%, respectively). This demonstrates its capability to reduce false positives with minimal loss in recall. By lowering misclassifications of non-discriminatory deeds, the pipeline reduces the number of documents routed to the Mapping Prejudice volunteers. As the project is a community endeavor in which volunteers review model-flagged documents, high precision directly eases their reading burden and improves the efficiency of downstream verification. 

However, our pipeline can still miss implicit restrictive clauses, such as ``No nationality, servants other different than race or nationality employed by an owner $\cdots$'', as it is trained to recognize explicit racial terms rather than make broader document-level inferences. A promising direction is a hybrid, two-stage approach: first, apply document-level classification to detect documents that likely contain racially restrictive content, then perform token-level sequence tagging to localize the terms. This cascaded design may reduce false negatives in implicit cases while preserving interpretability.


\subsection{Georeferencing Racial Deeds} \label{s:result-georeference}
For RPSS entities (i.e., State, County, City, and Subdivision), we evaluate the NER performance since the ground-truth labels are available. For PLSS entities (i.e., Township, Range, and Section), we assess the end-to-end georeferencing accuracy after resolving the NER results to the modern location of the deed documents.

For the RPSS entities, we report per-class precision, recall, and F1 scores (Table \ref{t:nerperformance}) using volunteer-labeled annotations curated by the Mapping Prejudice team as the ground truth. NER performance highly depends on the label quality. Subdivision labels, directly transcribed from deeds and reviewed multiple times, are high quality and, consequently, yield the highest F1 score (91.88\%). In contrast, State, County, and City labels exhibit substantial format variability due to OCR errors, abbreviations (e.g., `No.' used for `Number'), and historical changes in names and boundaries. Beyond these issues, City references are often incomplete in deed documents, which makes City the lowest-performing category out of the four labels.

\begin{table}[htbp]
    \centering \small
    \begin{tabular}{|c|c|c|c|}
        \hline
        \textbf{Entity} & \textbf{Precision (\%)} & \textbf{Recall (\%)} & \textbf{F1-score (\%)} \\
        \hline
        State       & 99.13 & 65.61 & 78.96 \\
        County      & 97.95 & 63.64 & 77.16 \\
        City        & 81.29 & 34.11 & 48.06 \\
        Subdivision & 95.74 & 88.31 & 91.88 \\
        \hline
    \end{tabular}
    \caption{Evaluation metrics of Entity Recognition in RPSS Labels.}
    \label{t:nerperformance}
\end{table}
For the PLSS entities, we assess the end-to-end georeferencing accuracy by checking whether volunteer-labeled deed property boundaries fall within or partially overlap the PLSS parcel boundaries that our pipeline identifies. We do not report separate NER metrics for PLSS since, unlike RPSS entities, we do not have token-level ground truth. As shown in Table \ref{t:georeferencing}, across 1,470 filtered deeds, we achieve 85.58\% accuracy at 1×1 square-mile resolution and 86.39\% at 6×6 square-mile resolution, indicating reliable georeferencing.

Subdivision-level georeferencing is valuable for studying racial covenants, as subdivisions are often smaller than 1×1 square miles and align with neighborhood-scale patterns. Our current subdivision-level georeferencing relies on fuzzy matching between NER outputs and modern administrative boundary names. Under these conditions, 16.18\% of deeds can reach subdivision-level georeferencing. The current NER training data annotation uses regular expressions for auto-matching, which cannot accurately cover the diverse variants of RPSS names in deeds. We see opportunities to expand this coverage and improve the recall of subdivision-level georeferencing through systematic normalization, learned matching, and the use of transformer-based models.


\begin{table}[htbp]
    \centering \small
    \begin{tabular}{c|c|c}
        \hline
        \textbf{Resolution} & \textbf{Entities} & \textbf{Accuracy} \\
        \hline
        \textbf{6×6 sq-mi} & Township, Range & 86.39\% (1,270 deeds) \\
        \textbf{1×1 sq-mi} & Township, Range, Section & 85.58\% (1,258 deeds) \\
        \hline
    \end{tabular}
    \caption{Georeferencing accuracy for PLSS entities at 1×1 sq-mi and 6×6 sq-mi resolutions.}
    \label{t:georeferencing}
\end{table}

\section{Conclusion}
Our results show that human-centered computing systems can strengthen public memory and historical accountability by improving the scale and depth of racial covenant mapping. Our racial entity identification pipeline reduces false positives by 25.96\% while maintaining a recall rate of 91.73\%. Additionally, our georeferencing pipeline can accurately suggest the contemporary location of the deed document within a 1 square-mile area with 85.58\% accuracy when evaluated across six counties. These advances improve the Mapping Prejudice workflow by streamlining repetitive steps, allowing volunteers to focus on interpretive tasks such as identifying implicit covenants and clarifying ambiguous geographic references. 

Human involvement remains crucial for catching inevitable errors from automation and ensuring the reliability of the results. In addition, we recognize that legal traditions, survey systems, and archival practice often differ across states, which means that our trained pipelines may not directly generalize beyond Minnesota and Wisconsin without additional adaptation. Given the complexities of historical and geographical contexts, acknowledging these differences highlights the importance of designing technology that supports, rather than replaces, human decision-making. This also opens opportunities for future work to refine and extend our approach in collaboration with communities in other regions, ensuring that the method remains sensitive to local context and historical specifications. Together, our efforts show that combining automation with civic participation can responsibly increase technical capacity while deepening community engagement by empowering community members to actively participate in uncovering their shared history.

\bibliographystyle{ACM-Reference-Format}
\bibliography{references}


\begin{thebibliography}{35}


\ifx \showCODEN    \undefined \def \showCODEN     #1{\unskip}     \fi
\ifx \showISBNx    \undefined \def \showISBNx     #1{\unskip}     \fi
\ifx \showISBNxiii \undefined \def \showISBNxiii  #1{\unskip}     \fi
\ifx \showISSN     \undefined \def \showISSN      #1{\unskip}     \fi
\ifx \showLCCN     \undefined \def \showLCCN      #1{\unskip}     \fi
\ifx \shownote     \undefined \def \shownote      #1{#1}          \fi
\ifx \showarticletitle \undefined \def \showarticletitle #1{#1}   \fi
\ifx \showURL      \undefined \def \showURL       {\relax}        \fi
\providecommand\bibfield[2]{#2}
\providecommand\bibinfo[2]{#2}
\providecommand\natexlab[1]{#1}
\providecommand\showeprint[2][]{arXiv:#2}

\bibitem[Aldana-Bobadilla et~al\mbox{.}(2020)]%
        {c:aldana2020}
\bibfield{author}{\bibinfo{person}{Edwin Aldana-Bobadilla}, \bibinfo{person}{Alejandro Molina-Villegas}, \bibinfo{person}{Ivan Lopez-Arevalo}, \bibinfo{person}{Shanel Reyes-Palacios}, \bibinfo{person}{Victor Muñiz-Sanchez}, {and} \bibinfo{person}{Jean Arreola-Trapala}.} \bibinfo{year}{2020}\natexlab{}.
\newblock \showarticletitle{Adaptive Geoparsing Method for Toponym Recognition and Resolution in Unstructured Text}.
\newblock \bibinfo{journal}{\emph{Remote Sensing}} \bibinfo{volume}{12}, \bibinfo{number}{18} (\bibinfo{year}{2020}).
\newblock
\showISSN{2072-4292}
\href{https://doi.org/10.3390/rs12183041}{doi:\nolinkurl{10.3390/rs12183041}}


\bibitem[Barriere and Cifuentes(2024)]%
        {c:barriere2024}
\bibfield{author}{\bibinfo{person}{Valentin Barriere} {and} \bibinfo{person}{Sebastian Cifuentes}.} \bibinfo{year}{2024}\natexlab{}.
\newblock \showarticletitle{A Study of Nationality Bias in Names and Perplexity using Off-the-Shelf Affect-related Tweet Classifiers}. In \bibinfo{booktitle}{\emph{Proceedings of the 2024 Conference on Empirical Methods in Natural Language Processing}}, \bibfield{editor}{\bibinfo{person}{Yaser Al-Onaizan}, \bibinfo{person}{Mohit Bansal}, {and} \bibinfo{person}{Yun-Nung Chen}} (Eds.). \bibinfo{publisher}{Association for Computational Linguistics}, \bibinfo{address}{Miami, Florida, USA}, \bibinfo{pages}{569--579}.
\newblock
\href{https://doi.org/10.18653/v1/2024.emnlp-main.34}{doi:\nolinkurl{10.18653/v1/2024.emnlp-main.34}}


\bibitem[Blodgett et~al\mbox{.}(2020)]%
        {c:blodgett2020}
\bibfield{author}{\bibinfo{person}{Su~Lin Blodgett}, \bibinfo{person}{Solon Barocas}, \bibinfo{person}{Hal Daum{\'e}~III}, {and} \bibinfo{person}{Hanna Wallach}.} \bibinfo{year}{2020}\natexlab{}.
\newblock \showarticletitle{Language (Technology) is Power: A Critical Survey of ``Bias'' in {NLP}}. In \bibinfo{booktitle}{\emph{Proceedings of the 58th Annual Meeting of the Association for Computational Linguistics}}, \bibfield{editor}{\bibinfo{person}{Dan Jurafsky}, \bibinfo{person}{Joyce Chai}, \bibinfo{person}{Natalie Schluter}, {and} \bibinfo{person}{Joel Tetreault}} (Eds.). \bibinfo{publisher}{Association for Computational Linguistics}, \bibinfo{address}{Online}, \bibinfo{pages}{5454--5476}.
\newblock
\href{https://doi.org/10.18653/v1/2020.acl-main.485}{doi:\nolinkurl{10.18653/v1/2020.acl-main.485}}


\bibitem[Bolukbasi et~al\mbox{.}(2016)]%
        {c:bolukbasi2016}
\bibfield{author}{\bibinfo{person}{Tolga Bolukbasi}, \bibinfo{person}{Kai-Wei Chang}, \bibinfo{person}{James Zou}, \bibinfo{person}{Venkatesh Saligrama}, {and} \bibinfo{person}{Adam Kalai}.} \bibinfo{year}{2016}\natexlab{}.
\newblock \showarticletitle{Man is to computer programmer as woman is to homemaker? debiasing word embeddings}. In \bibinfo{booktitle}{\emph{Proceedings of the 30th International Conference on Neural Information Processing Systems}} (Barcelona, Spain) \emph{(\bibinfo{series}{NIPS'16})}. \bibinfo{publisher}{Curran Associates Inc.}, \bibinfo{address}{Red Hook, NY, USA}, \bibinfo{pages}{4356–4364}.
\newblock
\showISBNx{9781510838819}


\bibitem[Clemence(2021)]%
        {c:clemence2021}
\bibfield{author}{\bibinfo{person}{Sara Clemence}.} \bibinfo{year}{2021}\natexlab{}.
\newblock \bibinfo{title}{Is There Racism in the Deed to Your Home?}
\newblock
\urldef\tempurl%
\url{https://www.nytimes.com/2021/08/17/realestate/racism-home-deeds.html}
\showURL{%
\tempurl}
\newblock
\shownote{Accessed 15-08-2025}.


\bibitem[Delegard and Corey(2024)]%
        {c:mappingprejudice}
\bibfield{author}{\bibinfo{person}{K. Delegard} {and} \bibinfo{person}{M. Corey}.} \bibinfo{year}{2024}\natexlab{}.
\newblock \bibinfo{booktitle}{\emph{Mapping Racial Covenants in the United States: A Technical Toolkit}}.
\newblock \bibinfo{type}{{T}echnical {R}eport}. \bibinfo{institution}{Univ. of Minnesota}.
\newblock
\newblock
\shownote{Report for the National Endowment for the Humanities}.


\bibitem[Devlin et~al\mbox{.}(2019)]%
        {c:devlin2019}
\bibfield{author}{\bibinfo{person}{Jacob Devlin}, \bibinfo{person}{Ming-Wei Chang}, \bibinfo{person}{Kenton Lee}, {and} \bibinfo{person}{Kristina Toutanova}.} \bibinfo{year}{2019}\natexlab{}.
\newblock \showarticletitle{{BERT}: Pre-training of Deep Bidirectional Transformers for Language Understanding}. In \bibinfo{booktitle}{\emph{Proceedings of the 2019 Conference of the North {A}merican Chapter of the Association for Computational Linguistics: Human Language Technologies, Volume 1 (Long and Short Papers)}}, \bibfield{editor}{\bibinfo{person}{Jill Burstein}, \bibinfo{person}{Christy Doran}, {and} \bibinfo{person}{Thamar Solorio}} (Eds.). \bibinfo{publisher}{Association for Computational Linguistics}, \bibinfo{address}{Minneapolis, Minnesota}, \bibinfo{pages}{4171--4186}.
\newblock
\href{https://doi.org/10.18653/v1/N19-1423}{doi:\nolinkurl{10.18653/v1/N19-1423}}


\bibitem[Ehrman-Solberg et~al\mbox{.}(2022)]%
        {c:ehrman2022}
\bibfield{author}{\bibinfo{person}{Kevin Ehrman-Solberg}, \bibinfo{person}{Bonnie Keeler}, \bibinfo{person}{Kate Derickson}, {and} \bibinfo{person}{Kirsten Delegard}.} \bibinfo{year}{2022}\natexlab{}.
\newblock \showarticletitle{Mapping a path towards equity: reflections on a co-creative community praxis}.
\newblock \bibinfo{journal}{\emph{GeoJournal}} \bibinfo{volume}{87}, \bibinfo{number}{S2} (\bibinfo{date}{Aug.} \bibinfo{year}{2022}), \bibinfo{pages}{185--194}.
\newblock


\bibitem[Goetz et~al\mbox{.}(2020)]%
        {c:goetz2024}
\bibfield{author}{\bibinfo{person}{E.~G. Goetz}, \bibinfo{person}{R.~A. Williams}, {and} \bibinfo{person}{A. Damiano}.} \bibinfo{year}{2020}\natexlab{}.
\newblock \showarticletitle{Whiteness and Urban Planning}.
\newblock \bibinfo{journal}{\emph{Journal of the American Planning Association}} \bibinfo{volume}{86}, \bibinfo{number}{2} (\bibinfo{year}{2020}), \bibinfo{pages}{142--156}.
\newblock


\bibitem[Gritta et~al\mbox{.}(2020)]%
        {c:gritta2020}
\bibfield{author}{\bibinfo{person}{Milan Gritta}, \bibinfo{person}{Mohammad~Taher Pilehvar}, {and} \bibinfo{person}{Nigel Collier}.} \bibinfo{year}{2020}\natexlab{}.
\newblock \showarticletitle{A pragmatic guide to geoparsing evaluation: Toponyms, Named Entity Recognition and pragmatics}.
\newblock \bibinfo{journal}{\emph{Lang. Resour. Eval.}} \bibinfo{volume}{54}, \bibinfo{number}{3} (\bibinfo{year}{2020}), \bibinfo{pages}{683--712}.
\newblock


\bibitem[Hu et~al\mbox{.}(2023)]%
        {c:hu2023}
\bibfield{author}{\bibinfo{person}{Xuke Hu}, \bibinfo{person}{Zhiyong Zhou}, \bibinfo{person}{Hao Li}, \bibinfo{person}{Yingjie Hu}, \bibinfo{person}{Fuqiang Gu}, \bibinfo{person}{Jens Kersten}, \bibinfo{person}{Hongchao Fan}, {and} \bibinfo{person}{Friederike Klan}.} \bibinfo{year}{2023}\natexlab{}.
\newblock \showarticletitle{Location Reference Recognition from Texts: A Survey and Comparison}.
\newblock \bibinfo{journal}{\emph{ACM Comput. Surv.}} \bibinfo{volume}{56}, \bibinfo{number}{5}, Article \bibinfo{articleno}{112} (\bibinfo{date}{Nov.} \bibinfo{year}{2023}), \bibinfo{numpages}{37}~pages.
\newblock
\showISSN{0360-0300}
\href{https://doi.org/10.1145/3625819}{doi:\nolinkurl{10.1145/3625819}}


\bibitem[Kahlenberg(2019)]%
        {c:kahlenberg2019}
\bibfield{author}{\bibinfo{person}{Richard Kahlenberg}.} \bibinfo{year}{2019}\natexlab{}.
\newblock \bibinfo{title}{How Minneapolis Ended Single-Family Zoning}.
\newblock
\urldef\tempurl%
\url{https://tcf.org/content/report/minneapolis-ended-single-family-zoning/}
\showURL{%
\tempurl}
\newblock
\shownote{Accessed 15-08-2025}.


\bibitem[Mamta et~al\mbox{.}(2024)]%
        {c:mamta2024}
\bibfield{author}{\bibinfo{person}{Mamta Mamta}, \bibinfo{person}{Rishikant Chigrupaatii}, {and} \bibinfo{person}{Asif Ekbal}.} \bibinfo{year}{2024}\natexlab{}.
\newblock \showarticletitle{{B}ias{W}ipe: Mitigating Unintended Bias in Text Classifiers through Model Interpretability}. In \bibinfo{booktitle}{\emph{Proceedings of the 2024 Conference on Empirical Methods in Natural Language Processing}}, \bibfield{editor}{\bibinfo{person}{Yaser Al-Onaizan}, \bibinfo{person}{Mohit Bansal}, {and} \bibinfo{person}{Yun-Nung Chen}} (Eds.). \bibinfo{publisher}{Association for Computational Linguistics}, \bibinfo{address}{Miami, Florida, USA}, \bibinfo{pages}{21059--21070}.
\newblock
\href{https://doi.org/10.18653/v1/2024.emnlp-main.1172}{doi:\nolinkurl{10.18653/v1/2024.emnlp-main.1172}}


\bibitem[Mattke et~al\mbox{.}(2022)]%
        {c:mattke2022}
\bibfield{author}{\bibinfo{person}{R. Mattke}, \bibinfo{person}{K. Delegard}, {and} \bibinfo{person}{D. Leebaw}.} \bibinfo{year}{2022}\natexlab{}.
\newblock \showarticletitle{Mapping Prejudice: The Map Library as a Hub for Community Co-Creation and Social Change}.
\newblock \bibinfo{journal}{\emph{J. Map Geogr. Libr.}} \bibinfo{volume}{18}, \bibinfo{number}{1-2} (\bibinfo{year}{2022}), \bibinfo{pages}{1--21}.
\newblock
\showISSN{1542-0353}


\bibitem[McCormick(2020)]%
        {c:mccormick2020}
\bibfield{author}{\bibinfo{person}{Kathleen McCormick}.} \bibinfo{year}{2020}\natexlab{}.
\newblock \bibinfo{title}{Rezoning History: Influential Minneapolis Policy Shift Links Affordability, Equity}.
\newblock
\urldef\tempurl%
\url{https://www.lincolninst.edu/publications/issues/land-lines-january-2020}
\showURL{%
\tempurl}
\newblock
\shownote{Accessed 15-08-2025}.


\bibitem[{Minnesota Natural Resources Department}(2023)]%
        {c:MN_PLSS_2023}
\bibfield{author}{\bibinfo{person}{{Minnesota Natural Resources Department}}.} \bibinfo{year}{2023}\natexlab{}.
\newblock \bibinfo{title}{Control Point Generated Public Land Survey (PLS)}.
\newblock \bibinfo{howpublished}{GIS Dataset}.
\newblock
\urldef\tempurl%
\url{https://gisdata.mn.gov/dataset/plan-mndnr-public-land-survey}
\showURL{%
\tempurl}
\newblock
\shownote{Accessed via Minnesota Geospatial Information Office (MnGeo) Data Portal; Spatial coverage: Minnesota, USA; Dataset provides PLSS-related geospatial data for public land survey applications.}.


\bibitem[Mogush and Worthington(2020)]%
        {c:Mogush02012020}
\bibfield{author}{\bibinfo{person}{Paul Mogush} {and} \bibinfo{person}{Heather Worthington}.} \bibinfo{year}{2020}\natexlab{}.
\newblock \showarticletitle{The View From Minneapolis: Comments on “Death to Single-Family Zoning” and “It’s Time to End Single-Family Zoning”}.
\newblock \bibinfo{journal}{\emph{Journal of the American Planning Association}} \bibinfo{volume}{86}, \bibinfo{number}{1} (\bibinfo{year}{2020}), \bibinfo{pages}{120--120}.
\newblock
\showeprint{https://doi.org/10.1080/01944363.2019.1689012}
\href{https://doi.org/10.1080/01944363.2019.1689012}{doi:\nolinkurl{10.1080/01944363.2019.1689012}}


\bibitem[Nadeem et~al\mbox{.}(2021)]%
        {c:nadeem2021}
\bibfield{author}{\bibinfo{person}{Moin Nadeem}, \bibinfo{person}{Anna Bethke}, {and} \bibinfo{person}{Siva Reddy}.} \bibinfo{year}{2021}\natexlab{}.
\newblock \showarticletitle{{S}tereo{S}et: Measuring stereotypical bias in pretrained language models}. In \bibinfo{booktitle}{\emph{Proceedings of the 59th Annual Meeting of the Association for Computational Linguistics and the 11th International Joint Conference on Natural Language Processing (Volume 1: Long Papers)}}, \bibfield{editor}{\bibinfo{person}{Chengqing Zong}, \bibinfo{person}{Fei Xia}, \bibinfo{person}{Wenjie Li}, {and} \bibinfo{person}{Roberto Navigli}} (Eds.). \bibinfo{publisher}{Association for Computational Linguistics}, \bibinfo{address}{Online}, \bibinfo{pages}{5356--5371}.
\newblock
\href{https://doi.org/10.18653/v1/2021.acl-long.416}{doi:\nolinkurl{10.18653/v1/2021.acl-long.416}}


\bibitem[Nangia et~al\mbox{.}(2020)]%
        {c:nangia2020}
\bibfield{author}{\bibinfo{person}{Nikita Nangia}, \bibinfo{person}{Clara Vania}, \bibinfo{person}{Rasika Bhalerao}, {and} \bibinfo{person}{Samuel~R. Bowman}.} \bibinfo{year}{2020}\natexlab{}.
\newblock \showarticletitle{{C}row{S}-Pairs: A Challenge Dataset for Measuring Social Biases in Masked Language Models}. In \bibinfo{booktitle}{\emph{Proceedings of the 2020 Conference on Empirical Methods in Natural Language Processing (EMNLP)}}, \bibfield{editor}{\bibinfo{person}{Bonnie Webber}, \bibinfo{person}{Trevor Cohn}, \bibinfo{person}{Yulan He}, {and} \bibinfo{person}{Yang Liu}} (Eds.). \bibinfo{publisher}{Association for Computational Linguistics}, \bibinfo{address}{Online}, \bibinfo{pages}{1953--1967}.
\newblock
\href{https://doi.org/10.18653/v1/2020.emnlp-main.154}{doi:\nolinkurl{10.18653/v1/2020.emnlp-main.154}}


\bibitem[of~Minneapolis Department~of Community~Planning and Development(2019)]%
        {c:minneapolis2040}
\bibfield{author}{\bibinfo{person}{City of~Minneapolis Department~of Community~Planning} {and} \bibinfo{person}{Economic Development}.} \bibinfo{year}{2019}\natexlab{}.
\newblock \bibinfo{title}{Minneapolis 2040}.
\newblock
\urldef\tempurl%
\url{https://minneapolis2040.com/}
\showURL{%
\tempurl}
\newblock
\shownote{Accessed August 15, 2025}.


\bibitem[Qian et~al\mbox{.}(2019)]%
        {c:qian2019}
\bibfield{author}{\bibinfo{person}{Yusu Qian}, \bibinfo{person}{Urwa Muaz}, \bibinfo{person}{Ben Zhang}, {and} \bibinfo{person}{Jae~Won Hyun}.} \bibinfo{year}{2019}\natexlab{}.
\newblock \showarticletitle{Reducing Gender Bias in Word-Level Language Models with a Gender-Equalizing Loss Function}. In \bibinfo{booktitle}{\emph{Proceedings of the 57th Annual Meeting of the Association for Computational Linguistics: Student Research Workshop}}, \bibfield{editor}{\bibinfo{person}{Fernando Alva-Manchego}, \bibinfo{person}{Eunsol Choi}, {and} \bibinfo{person}{Daniel Khashabi}} (Eds.). \bibinfo{publisher}{Association for Computational Linguistics}, \bibinfo{address}{Florence, Italy}, \bibinfo{pages}{223--228}.
\newblock
\href{https://doi.org/10.18653/v1/P19-2031}{doi:\nolinkurl{10.18653/v1/P19-2031}}


\bibitem[Santucci(2025)]%
        {c:santucci2025}
\bibfield{author}{\bibinfo{person}{L. Santucci}.} \bibinfo{year}{2025}\natexlab{}.
\newblock \bibinfo{booktitle}{\emph{The racial wealth gap and the legacy of racially restrictive housing covenants}}.
\newblock \bibinfo{publisher}{Oxford Univ. Press}. 135--152 pages.
\newblock


\bibitem[Sood et~al\mbox{.}(2019)]%
        {c:sood2019}
\bibfield{author}{\bibinfo{person}{A. Sood}, \bibinfo{person}{W. Speagle}, {and} \bibinfo{person}{K. Ehrman-Solberg}.} \bibinfo{year}{2019}\natexlab{}.
\newblock \bibinfo{booktitle}{\emph{Long Shadow of Racial Discrimination: Evidence from Housing Covenants of Minneapolis}}.
\newblock \bibinfo{type}{Working Paper}.
\newblock


\bibitem[Staff(2012)]%
        {c:plss2012}
\bibfield{author}{\bibinfo{person}{Staff}.} \bibinfo{year}{2012}\natexlab{}.
\newblock \bibinfo{title}{The Public Land Survey System (PLSS)}.
\newblock
\urldef\tempurl%
\url{https://web.archive.org/web/20120607063232/http://www.nationalatlas.gov/articles/boundaries/a_plss.html}
\showURL{%
\tempurl}
\newblock
\shownote{Archived June 7, 2012; Retrieved July 31, 2025}.


\bibitem[Sun et~al\mbox{.}(2019)]%
        {c:sun2019}
\bibfield{author}{\bibinfo{person}{Tony Sun}, \bibinfo{person}{Andrew Gaut}, \bibinfo{person}{Shirlyn Tang}, \bibinfo{person}{Yuxin Huang}, \bibinfo{person}{Mai ElSherief}, \bibinfo{person}{Jieyu Zhao}, \bibinfo{person}{Diba Mirza}, \bibinfo{person}{Elizabeth Belding}, \bibinfo{person}{Kai-Wei Chang}, {and} \bibinfo{person}{William~Yang Wang}.} \bibinfo{year}{2019}\natexlab{}.
\newblock \showarticletitle{Mitigating Gender Bias in Natural Language Processing: Literature Review}. In \bibinfo{booktitle}{\emph{Proceedings of the 57th Annual Meeting of the Association for Computational Linguistics}}, \bibfield{editor}{\bibinfo{person}{Anna Korhonen}, \bibinfo{person}{David Traum}, {and} \bibinfo{person}{Llu{\'i}s M{\`a}rquez}} (Eds.). \bibinfo{publisher}{Association for Computational Linguistics}, \bibinfo{address}{Florence, Italy}, \bibinfo{pages}{1630--1640}.
\newblock
\href{https://doi.org/10.18653/v1/P19-1159}{doi:\nolinkurl{10.18653/v1/P19-1159}}


\bibitem[{Supreme Court of the United States}(1948)]%
        {c:shelleyvkraemer}
\bibfield{author}{\bibinfo{person}{{Supreme Court of the United States}}.} \bibinfo{year}{1948}\natexlab{}.
\newblock \bibinfo{title}{Shelley v. Kraemer, 334 U.S. 1 (1948)}.
\newblock
\newblock
\shownote{U.S. Reports: Shelley v. Kraemer, 334 U.S. 1}.


\bibitem[{Thomson Reuters}(2025)]%
        {c:reuterslot2025}
\bibfield{author}{\bibinfo{person}{{Thomson Reuters}}.} \bibinfo{year}{2025}\natexlab{}.
\newblock \bibinfo{title}{Lot and Block Survey System}.
\newblock
\urldef\tempurl%
\url{https://content.next.westlaw.com/5-507-1532}
\showURL{%
\tempurl}
\newblock
\shownote{In \textit{Practical Law Glossary}, Westlaw. Retrieved July 31, 2025}.


\bibitem[(USGS)(2025)]%
        {c:usgs2025plss}
\bibfield{author}{\bibinfo{person}{U.S. Geol.~Surv. (USGS)}.} \bibinfo{year}{2025}\natexlab{}.
\newblock \bibinfo{title}{Do US Topos and The National Map have a layer that shows the Public Land Survey System (PLSS)?}
\newblock
\urldef\tempurl%
\url{https://www.usgs.gov/faqs/do-us-topos-and-national-map-have-a-layer-shows-public-land-survey-system-plss}
\showURL{%
\tempurl}
\newblock
\shownote{Updated Jan. 29, 2025; Accessed July 31, 2025}.


\bibitem[Walker and Derickson(2023)]%
        {c:walker2023}
\bibfield{author}{\bibinfo{person}{R.~H. Walker} {and} \bibinfo{person}{K.~D. Derickson}.} \bibinfo{year}{2023}\natexlab{}.
\newblock \showarticletitle{Mapping Prejudice}.
\newblock \bibinfo{journal}{\emph{Journal of the American Planning Association}} \bibinfo{volume}{89}, \bibinfo{number}{4} (\bibinfo{year}{2023}), \bibinfo{pages}{459--471}.
\newblock


\bibitem[Walker et~al\mbox{.}(2024)]%
        {c:walker2024}
\bibfield{author}{\bibinfo{person}{R.~H. Walker}, \bibinfo{person}{B.~L. Keeler}, {and} \bibinfo{person}{K.~D. Derickson}.} \bibinfo{year}{2024}\natexlab{}.
\newblock \showarticletitle{The impacts of racially discriminatory housing policies on the distribution of intra-urban heat and tree canopy: A comparison of racial covenants and redlining in Minneapolis, MN}.
\newblock \bibinfo{journal}{\emph{Landscape Urban Plann.}}  \bibinfo{volume}{245} (\bibinfo{year}{2024}), \bibinfo{pages}{105019}.
\newblock
\showISSN{0169-2046}


\bibitem[Warner et~al\mbox{.}(2024)]%
        {c:warner2024}
\bibfield{author}{\bibinfo{person}{B. Warner}, \bibinfo{person}{A. Chaffin}, \bibinfo{person}{B. Clavié}, \bibinfo{person}{O. Weller}, \bibinfo{person}{O. Hallström}, \bibinfo{person}{S. Taghadouini}, \bibinfo{person}{A. Gallagher}, \bibinfo{person}{R. Biswas}, \bibinfo{person}{F. Ladhak}, \bibinfo{person}{T. Aarsen}, \bibinfo{person}{N. Cooper}, \bibinfo{person}{G. Adams}, \bibinfo{person}{J. Howard}, {and} \bibinfo{person}{I. Poli}.} \bibinfo{year}{2024}\natexlab{}.
\newblock \bibinfo{title}{Smarter, Better, Faster, Longer: A Modern Bidirectional Encoder for Fast, Memory Efficient, and Long Context Finetuning and Inference}.
\newblock


\bibitem[Welsh(2018)]%
        {c:welsh2018}
\bibfield{author}{\bibinfo{person}{N.~H. Welsh}.} \bibinfo{year}{2018}\natexlab{}.
\newblock \showarticletitle{Racially Restrictive Covenants in the United States: A Call to Action}.
\newblock \bibinfo{journal}{\emph{Agora J. Urban Plann. Des.}} (\bibinfo{year}{2018}), \bibinfo{pages}{130--142}.
\newblock


\bibitem[{Wisconsin Department of Natural Resources (WI DNR)}(2019)]%
        {c:WI_PLSS_2019}
\bibfield{author}{\bibinfo{person}{{Wisconsin Department of Natural Resources (WI DNR)}}.} \bibinfo{year}{2019}\natexlab{}.
\newblock \bibinfo{title}{PLSS Sections}.
\newblock \bibinfo{howpublished}{GIS Dataset}.
\newblock
\urldef\tempurl%
\url{https://data-wi-dnr.opendata.arcgis.com/datasets/wi-dnr::plss-sections/explore?location=43.073405%2C-88.003446%2C9.66}
\showURL{%
\tempurl}
\newblock
\shownote{Accessed via ArcGIS Open Data; Spatial coverage: Wisconsin, USA; Dataset focuses on Public Land Survey System (PLSS) section boundaries.}.


\bibitem[Worthington(2020)]%
        {c:Worthington02012020}
\bibfield{author}{\bibinfo{person}{Heather Worthington}.} \bibinfo{year}{2020}\natexlab{}.
\newblock \showarticletitle{The Minneapolis 2040 Comprehensive Plan: Community Engagement and Policy Development Addressing Housing}.
\newblock \bibinfo{journal}{\emph{Technology|Architecture + Design}} \bibinfo{volume}{4}, \bibinfo{number}{1} (\bibinfo{year}{2020}), \bibinfo{pages}{120--123}.
\newblock
\showeprint{https://doi.org/10.1080/24751448.2020.1705735}
\href{https://doi.org/10.1080/24751448.2020.1705735}{doi:\nolinkurl{10.1080/24751448.2020.1705735}}


\bibitem[Zhang et~al\mbox{.}(2018)]%
        {c:zhang2018}
\bibfield{author}{\bibinfo{person}{Brian~Hu Zhang}, \bibinfo{person}{Blake Lemoine}, {and} \bibinfo{person}{Margaret Mitchell}.} \bibinfo{year}{2018}\natexlab{}.
\newblock \showarticletitle{Mitigating Unwanted Biases with Adversarial Learning}. In \bibinfo{booktitle}{\emph{Proceedings of the 2018 AAAI/ACM Conference on AI, Ethics, and Society}} (New Orleans, LA, USA) \emph{(\bibinfo{series}{AIES '18})}. \bibinfo{publisher}{Association for Computing Machinery}, \bibinfo{address}{New York, NY, USA}, \bibinfo{pages}{335–340}.
\newblock
\showISBNx{9781450360128}
\href{https://doi.org/10.1145/3278721.3278779}{doi:\nolinkurl{10.1145/3278721.3278779}}


\end{thebibliography}
\theendnotes

\end{document}